\documentclass[%
 pre,
 amsmath,amssymb,
 reprint,%
]{revtex4-1}

\usepackage{graphicx}
\usepackage{dcolumn}
\usepackage{bm}
\usepackage[utf8]{inputenc}
\usepackage[T1]{fontenc}
\usepackage{mathptmx}

\begin{document}


\title[Optical generation of plasma electromagnetic guiding structures with PW laser]{Optical generation of quasi-stationary plasma electromagnetic structures for particle collimation with PetaWatt picosecond laser}

\author{Ph.~Korneev}
 \altaffiliation[Also at ]{Lebedev Physical Institute, Moscow, Russian Federation.}
 \email{ph.korneev@gmail.com}
\author{N.~D.~Bukharskii}%
\author{I.~V.~Kochetkov}%
\affiliation{ 
National Research Nuclear University MEPhI, Moscow, Russian Federation
}%
\author{M.~Ehret}%
 \altaffiliation[Also at ]{Institut für Kernphysik, Technische Universität Darmstadt, Germany.}
\author{J.~J.~Santos}%
\affiliation{%
Laboratoire CELIA, Université de Bordeaux, France
}%

\author{Y.~Abe}%
\affiliation{%
Institute of Laser Engineering, Osaka University, Japan
}%
\author{K.~F.~F.~Law}%
\affiliation{%
Department of Earth and Planetary Science, Graduate School of Science, The University of Tokyo, Japan
}
\author{S.~Fujioka}%
\affiliation{%
Institute of Laser Engineering, Osaka University, Japan
}%

\author{G.~Schaumann}
\affiliation{Institut für Kernphysik, Technische Universität Darmstadt, Germany}

\author{B.~Zielbauer}%
\affiliation{Plasma Physik/PHELIX, GSI Helmholtzzentrum f\"{u}r Schwerionenforschung GmbH, Darmstadt,
Germany}

\date{\today}

\begin{abstract}
Corresponding author: Ph. Korneev, ph.korneev@gmail.com\\
Optical generation of energetic particle bunches requires high-power laser facilities operating in picosecond or femtosecond temporal domain. It is therefore preferable to use short laser pulses in all-optical platforms designed for guiding and focusing of such particle beams, increasing their brightness and decreasing their angular divergence. We propose and discuss theoretical and experimental results for a novel electromagnetic guiding setup based on a shaped spiral-like 'snail' target with a relatively large useful aperture. Due to the diameter increased to a sub-mm scale, a wider particle bunches may be efficiently focused, as we show in theoretical modelling with a model field distributions, supported by the experimental data.   
\end{abstract}

\maketitle



\section{\label{sec:Intro}Introduction}

Modern high-power laser facilities advanced plenty of applications based on accelerated particle beams. 
Along with expectations of great prospects in energy of accelerated protons~\cite{Robson2007}, fast advances are achieved in proton acceleration from nanostructured targets~\cite{Lubcke2017, Thiele2019, Margarone2012} and from microchannels~\cite{Zou2017}.
The quality of particle beams from conventional accelerators is different than that from laser-based schemes, though the compactness and total charge in the beams for the latter may be much higher. Among many propositions for particle beam collimation, applying either auxiliary guiding quasi-stationary electro-magnetic field~\cite{Bailly-Grandvaux2016,Chen-pop14} or spontaneous field structure, generated within the irradiated target~\cite{robinson-pop07,Robinson2008,Malko2019,Korneev-arxiv17} are thought to be one of the most efficient solutions. For desired particle beam characteristic, the conventional electromagnets fed by a capacitor discharge bank may appear to be either inadequate or inconvenient, however, optical generators, driven with powerful lasers, can provide sufficiently strong fields and do not need a cumbersome damage protection as they are normally cheep and single-used, as well as the targets for particle generation themselves.       

Various approaches to strong magnetic field generation with powerful lasers have been explored in the literature~\cite{Morita2023}. One of the most effective and widely used mechanism for laser excitation of magnetic field is the generation of return, or discharge current in a given geometry, like a single-turn coil~\cite{Korobkin-stpl79, daido-prl86, Fujioka2013a, Santos-NJP2015, Santos-pop2018}. To generate a sufficient potential, an interaction region may have a capacitor-like design, though the optimal parameters are still widely discussed. Those capacitor-coil targets have a few-mm spatial scale and are well characterized in experimental studies while driven by a nanosecond terawatt laser pulse~\cite{Santos-NJP2015, GaoPhysPlasmas2016, Santos-pop2018, ChienPhysPlasmas2021, PeeblesPhysPlasmas2020, PeeblesPhysPlasmas2022, CourtoisJApplPhys2005, Bradford2020}. For short, picosecond-scale and more intense drivers, the large spatial scales lead to a complicated transient behavior like a discharge wave propagation~\cite{Kar-nc2016, ehret_guided_2023}, though at later times a quasi-stationary electromagnetic structure may form~\cite{ehret_guided_2023}. Though some attempts at strong magnetic field generation were reported for coil targets driven by a short, even fs laser pulse, the magnetic field there did not exceed the value of a few tens of tesla~\cite{Wang2018, ZhuAPL2018}, limiting the use of such systems for collimation of high energy particle flows. However, it is the short pulses that are normally used for particle acceleration, and the auxiliary electromagnetic field generators working with short laser pulses may appear to be more convenient, as they can use then e.g. a part of the same driver.  

Recently, a novel approach for generation of quasi-stationary electromagnetic structures, was proposed~\cite{Korneev-pre15}. It is in general based on a "whispering gallery effect"~\cite{Abe2018a} and was experimentally proved to work robustly both for long~\cite{Pisarczyk-scirep2018} and short laser pulses~\cite{ehret_kilotesla_2022,Law2020}, though for more intense picosecond drivers, the magnetic field appear to be much higher and reaches values of several kilotesla~\cite{Law2020}. Such high magnetic field magnitudes have previously been reached only in magnetic flux compression setups where a seed magnetic field is created in a cylindrical target and is then compressed via implosion driven by a powerful laser~\cite{GotchevPRL2009, KnauerPhysPlasmas2010, HohenbergerPhysPlasmas2012}. The scheme considered in this work is more versatile as it does not require an external high-energy driver for obtaining $\gtrsim 1$~kT magnetic field. The interaction region there is itself a curved ring-like surface, working as a current circuit in a magnetic field generation process. Due to the fact that the laser-surface interaction and magnetic field generation occur in the same volume, the main feature of this scheme is that it produces finally a magnetized plasma inside the cylinder-like target volume instead of the electric current bound to a conducting rod. The plasma is confined with inertia of the target material, providing a life-time of the generated fields about a hundred of picoseconds~\cite{ehret_kilotesla_2022}. Depending on the geometry of interaction and laser pulse parameters, the magnetized plasma structure may have different characteristics~\cite{Korneev-jophcs2017} and allows also to study fundamental processes in magnetized plasma themselves, e.g. relativistic reconnection similar to those near astrophysical X-ray sources~\cite{Law2020}. Some interesting applications may be based on the described phenomenon, like e.g. hot plasma confinement~\cite{Guskov2016} or control of particle beams, which is discussed in this work.  

The paper is organized as follows. First, we present results of the experimental study and their interpretation for a picosecond laser pulse interaction with curved targets at \mbox{PHELIX} laser facility at GSI, Darmstadt. We consider two types of targets: the "snail"-type target with the entrance in its upper parts, i.e. opposite to the stalk, which we call here "S6" target, and the one with the entrance in the lower part, which we call here "S9". We discuss the electromagnetic structure which is formed inside the target volume of the sub-millimeter scale. Finally, before conclusion, we consider theoretically a possibility of focusing fast bunches of protons and relativistic electrons with the discussed electromagnetic structure. 

\section{Experimental study}
\begin{figure}
    \centering
    \includegraphics[width=0.49\linewidth]{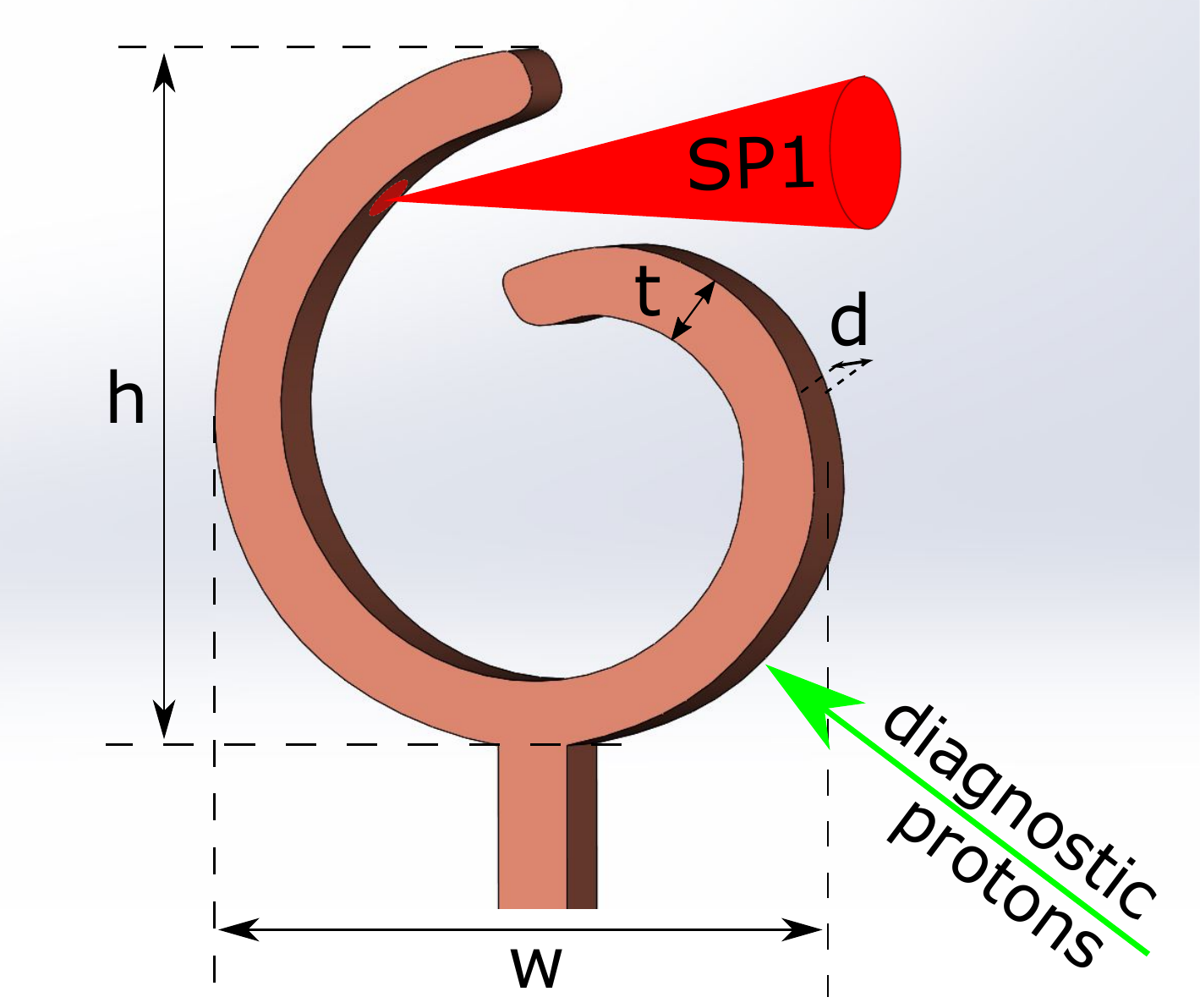}
    \hfil
    \includegraphics[width=0.46\linewidth]{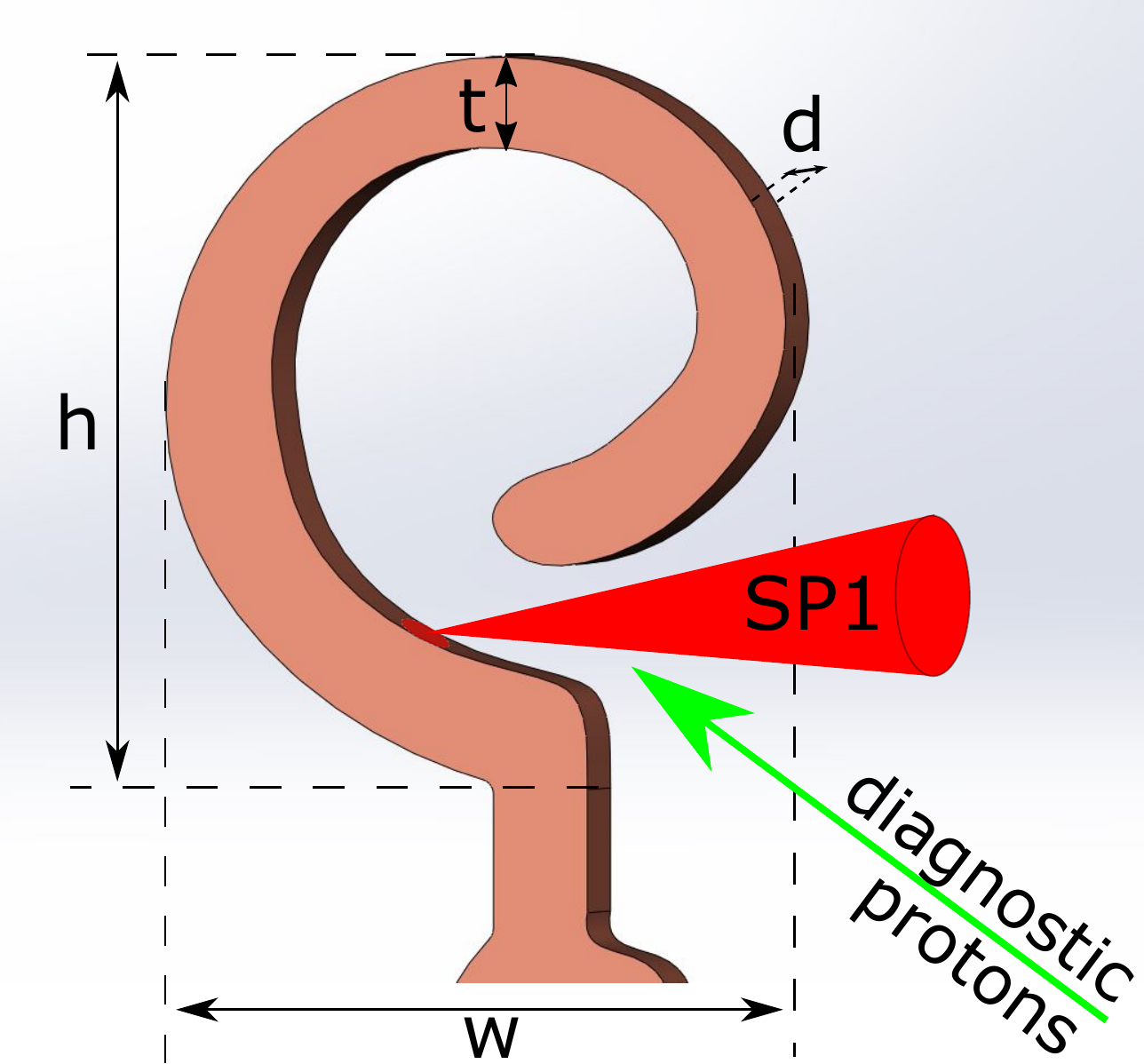}
    \caption{S6 (left) and S9 (right) targets with their sizes, SP1 beam and probing proton direction are shown schematically.}
    \label{fig:target_shapes}
\end{figure}

The experiment was carried out at the PHELIX laser facility at GSI, Darmstadt. The targets were produced at the Technical University Darmstadt using the MEMS (Micro-Electro-Mechanical Systems) technology. Their shapes and dimensions are presented in Fig.~\ref{fig:target_shapes} The target height (without the stalk) was $h=482~\mu$m, and the width was $w=424~\mu$m. Wall thickness was $\delta=50~\mu$m and the depth, i.e. dimension along the target axis, was $d=100 ~\mu$m. 

Targets used in the experiment were designed with the stalk positioned in two different ways, as explained above, for the S6 target and the S9 target, see Fig.~\ref{fig:target_shapes}. The S6 target photographic image is shown in the inset panel of Fig.~\ref{fig:exp_setup}. Using the two oppositely screwed targets within the same experiment allows to distinguish an input of the magnetic field which is expected to have opposite polarities in S6 and S9 targets. The targets used in the PIC simulations have the same 2D geometry as those used in the experiment.

\subsection{Experimental setup}
\begin{figure}
    \centering
    \includegraphics[width=0.95\linewidth]{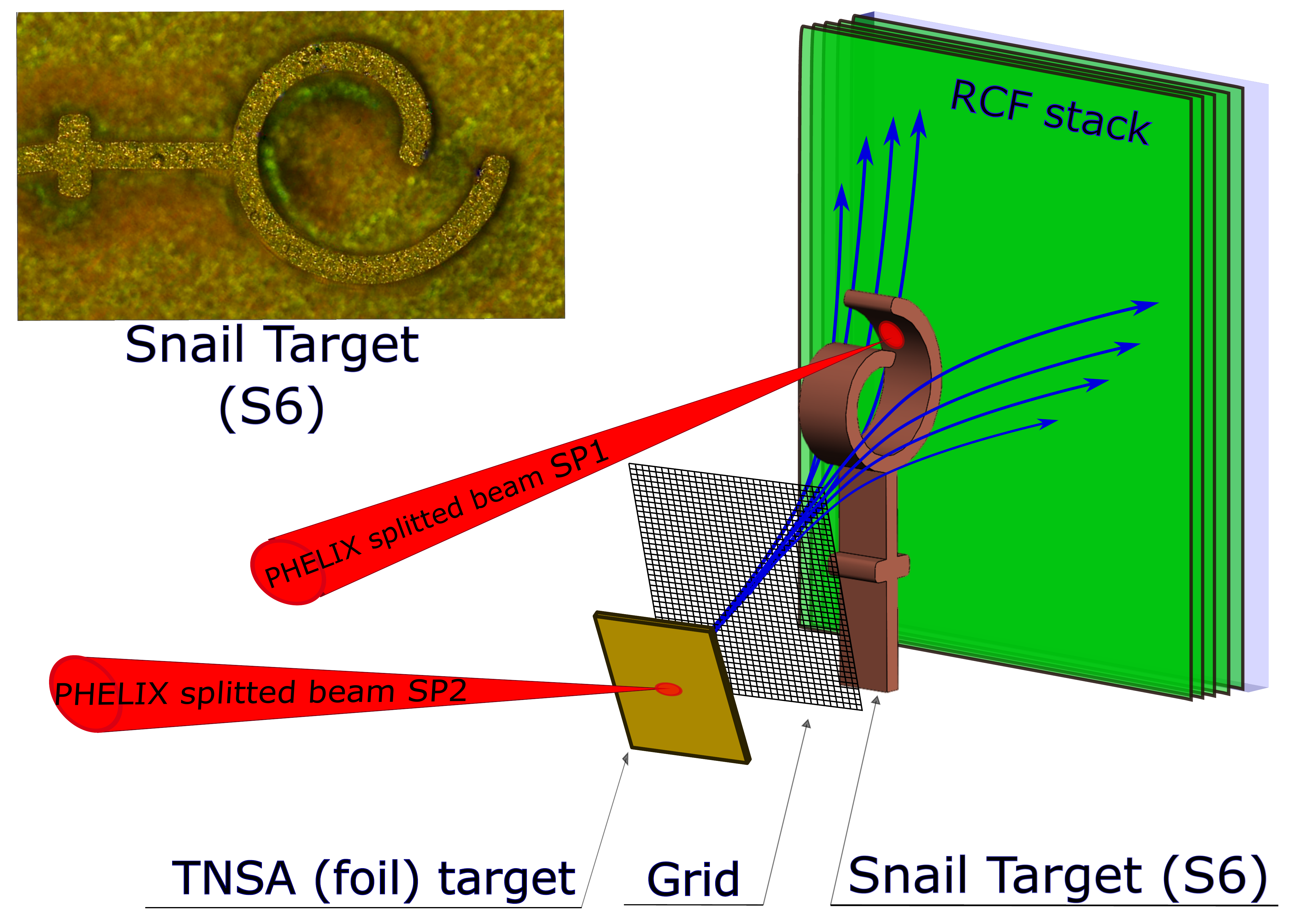}
    \caption{A simplified scheme of the experimental setup. The photographic image of the S6 “snail” target is shown aside.}
    \label{fig:exp_setup}
\end{figure}

 In the experiment, the laser pulse with 0.5~ps duration, 1.056~$\mu$m central wavelength and a high contrast~$\sim10^{-10}$, was split into two equal parts named SP1 and SP2, each carrying about 50~J. The SP1 beam was focused by a shorter focal length parabola $F=400$~mm into a $10~\mu$m FWHM (Full Width at Half Maximum) spot on the internal surface of the targets, as shown in Fig.~\ref{fig:target_shapes}. The expected intensity in the focal spot was a few times of $10^{19}$~W/cm$^{2}$. The SP2 beam was focused by a long focal length parabola $F=1500$~mm into a $40~\mu$m FWHM focal spot on a thin $4~\mu$m gold foil for generation of diagnostic protons via TNSA (Target Normal Sheath Acceleration) mechanism \cite{Wilks2001}. A simplified scheme of the full experimental setup is presented in Fig.~\ref{fig:exp_setup}. 

The proton deflectometry driven by the SP2 beam was a primary diagnostic in this experimental research. Multi-layer Radiochromic Film (RCF) stack was used as a proton detector. The stacks consisted of 15 $\mu$m Al foil to stop heavier particles, 5 HD-V2 RCFs (having a low sensitivity) and 10 EBT3 RCFs (having a high sensitivity). These RCFs can be used for quantitative dose retrieval using the existing calibration curves~\cite{Feng2018}. With different proton propagation till the Bragg absorption peak in stacked RCFs, it is possible to obtain a temporally resolved evolution of the fields in the studied region~\cite{Santos-NJP2015}.

The distance between the TNSA source and the target in the experiment was $2.8$~mm, while the distance from the target to the front of $5\times5$~cm RCF stack was $111$~mm, leading to magnification factor of $\approx40$. An additional adjustable delay of $\pm~500$~ps between SP1 and SP2 pulses allowed to study the field time dependence from the early moments of the interaction, taking into account the proton time of flight from the TNSA foil to the target.  A $1500/$inch mesh was placed between the foil and the target at $1.47$~mm distance from the foil for two reasons: (i) confirmation, that protons are originated in the foil and not in the target itself and (ii) as a possible additional tool for the field evaluation. The maximum registered proton energy reached $20$~MeV, though layers containing both the target shadow, modified by the fields and distinguishable mesh imprint are observed only for proton energies below $10$~MeV.

\subsection{Experimental results}
\begin{figure}
    \centering
    \includegraphics[width=0.95\linewidth]{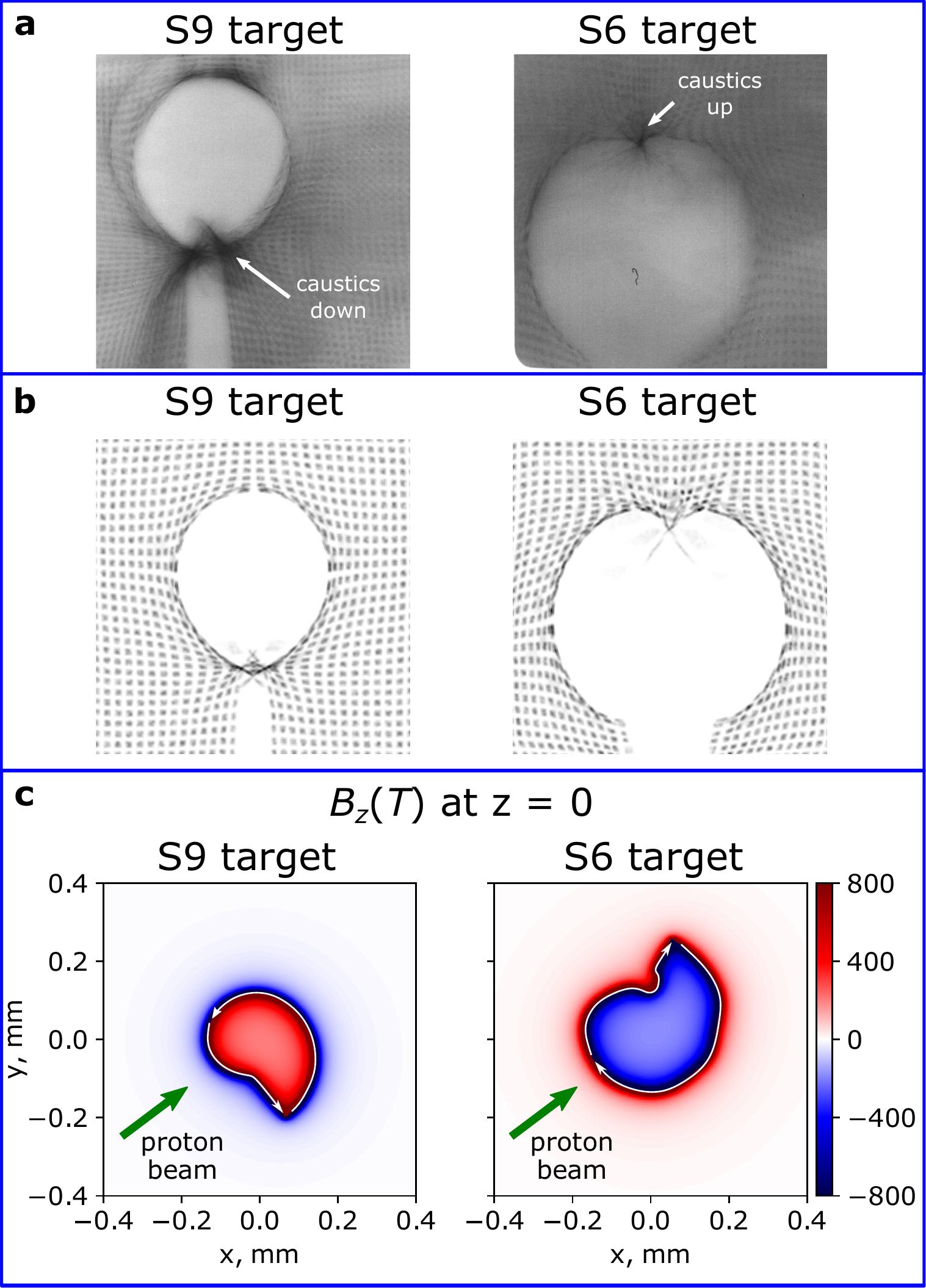}
    \caption{\textbf{a}:~Experimental proton radiography images, obtained $28\pm8$~ps after the laser pulse enters the S9 target  and $28\pm11$~ps after the laser pulse enters the S6 target. \textbf{b}:~Synthetic proton radiography images, obtained for the parameters of electromagnetic fields, deduced from the corresponding experimental images under the assumption of the homogeneous magnetic field structure. \textbf{c}:~Magnetic field profiles, used to estimate the average magnetic field in the target volume for both types of targets. The profiles are produced by a single current loop shown with a white line.}
    \label{fig:radiographs_and_MF_structure}
\end{figure}

In the experiment, several informative RCF data sets were obtained for both types of the targets, S6 and S9. Two exemplary radiograph images are shown in Fig.~\ref{fig:radiographs_and_MF_structure}, a. In each case the structure, observed on the radiograph, presents a region with a low proton signal. The boundary of this region has a cardioid shape with a distinct caustic feature either on the upper or on the lower side of the main 'void', depending on the polarity of the magnetic field, as we show below. Overall, these results are qualitatively similar to those observed for smaller targets~\cite{ehret_kilotesla_2022}. 

In order to analyse the radiographs and estimate the formed electro-magnetic fields, the test-particle approach was implemented. It requires calculating proton trajectories in different model electro-magnetic field distributions via numerical integration of the equations of motion. In this case, particles with energies matching those that produce an imprint on the corresponding layer of the RCF stack, were initialized at the position of the TNSA foil and sent through the studied region. Their trajectories in this region were calculated by solving the Newton's equations of motion with the Lorentz force $\vec F=e(\vec E+[\vec v\times \vec B])$, with a time step $0.2$~ps. Protons with trajectories intersecting the volume of the target material were removed from the simulation to account for the shadow of the target. Final coordinates of probe particles at the RCF plane formed a two-dimensional array, which was used for the synthetic proton radiography image construction. Electric fields in the simulations were calculated for a conducting target with a real geometry, charged to a certain potential, on a three-dimensional grid with $10~\mu$m resolution. Magnetic fields were calculated on the same grid by defining a current loop or several loops with certain electric currents, approximated by 100 linear segments and calculated then with the Biot-Savart law. For the initial analysis of the experimental data, the current loop was a curve that lies on the inner surface of the target and a straight segment that closed it in the slit region. In this case the magnetic field is distributed quite regularly inside the target cavity with a certain direction, see Fig.~\ref{fig:radiographs_and_MF_structure},~\textbf{c}. This simple circuit shape provides an estimation of the average magnetic field magnitude in the cavity. 

\begin{figure}
    \centering
    \includegraphics[width=0.95\linewidth]{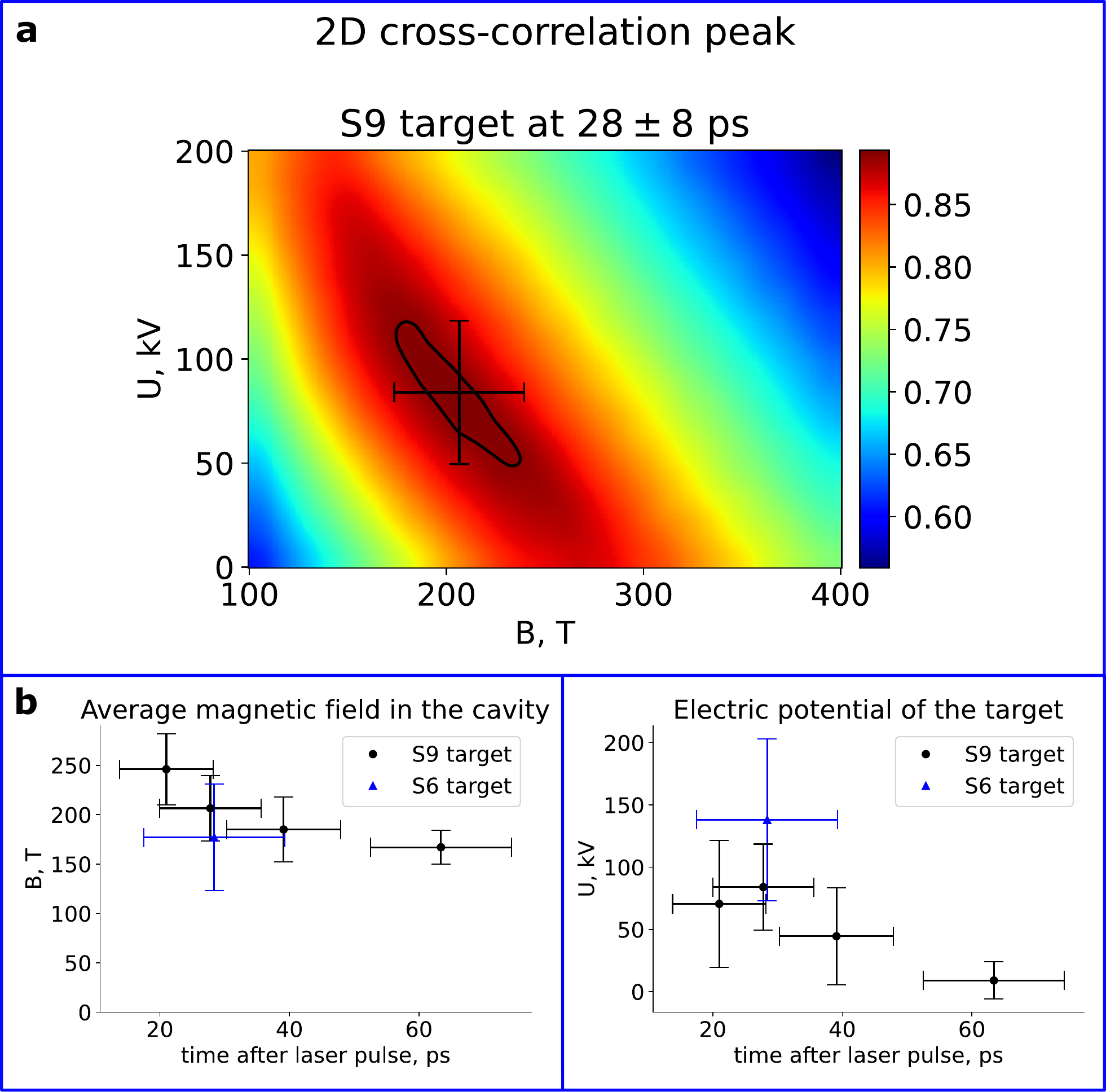}
    \caption{\textbf{a}:~Dependence of the peak value of cross-correlation between experimental and synthetic images on the average magnetic field inside the cavity and the electric potential of the target; the data corresponds to S9 target, probed $28 \pm 8$ ps after the laser pulse. \textbf{b}:~Temporal evolution of the average magnetic field inside the target cavity (left) and the electric potential of the target (right).}
    \label{fig:correlation_peak_and_B(t)}
\end{figure}

With the described "single current loop" approach, a set of synthetic radiographs for a range of different values of the target electric potential and electric current was obtained. 
The images from this data set were then compared to the experimental images by calculating a two-dimensional cross-correlation function. For each image from the set and the analyzed experimental image the maximum value of cross-correlation was determined. In this way, two-dimensional cross-correlation peak maps were built. For certain values of parameters, the correlation between experimental and synthetic image is noticeably higher, which can be interpreted as the best fit to experimental values within the considered model. Examples of synthetic images, calculated with the best-fit parameters for radiographs, presented in Fig.~\ref{fig:radiographs_and_MF_structure} \textbf{a}, are shown in Fig.~\ref{fig:radiographs_and_MF_structure} \textbf{b}. Since the two-dimensional correlation peak is not sharp and is elongated in the direction of the $UB\sim \mathrm{Const}$ line, where $U$ is electrostatic potential, $B$ is the magnetic field value, the best-fit region has a finite width along the both axis, resulting in uncertainties of extracted electromagnetic field parameters. An example of the two-dimensional correlation map, obtained for the experimental data for the S9 target, probed at $28\pm8$~ps after the laser pulse entrance the target, is demonstrated in Fig.~\ref{fig:correlation_peak_and_B(t)} \textbf{a}, where the best-fit region is enclosed in a black contour. Similar maps were calculated for other informative diagnostic images, and the result, namely the temporal evolution of the average magnetic field inside the cavity as well as the temporal evolution of the electric potential of the target, is plotted as a function of time after the laser pulse arrival in Fig.~\ref{fig:correlation_peak_and_B(t)} \textbf{b}. From the plot it can be seen that the magnetic field inside the S9 target is $250 \pm 40$~T at $21 \pm 7$~ps, and then it decays to $167 \pm 17$~T during the next $\approx~40$~ps. Some of the informative data for the S6 target corresponds to earlier moments of the field temporal evolution, when it is not yet formed, thus making its analysis in the assumption of a single current loop with a fixed shape inapplicable. For this reason, only one data point for the latest available time moment is shown for this target. The average field in this case is slightly lower than for the S9 target at the same moment of time, which can be attributed to the field distribution over a larger area inside the target, see Fig.~\ref{fig:radiographs_and_MF_structure} \textbf{b}, or to the worse interaction conditions, e.g. poor focusing on the inner surface of the target, resulting in a lack of multiple reflections or lower laser intensity on the surface. The electric potential for the S9 target decays faster than the magnetic field, changing from the initial $70 \pm 50$~kV to almost zero level at $63 \pm 11$~ps. It is worth noting that extraction errors for the electric potential are quite high, since it affects the radiochromic image less than the magnetic field, resulting in high correlation peak width along the $U$-axis.

As it is discussed below, the "single current loop" approach is not the only possible current distribution which may be attributed to the experimental data. However, the reconstructed values of the magnetic fields correctly describe both the average (integrated along the probing protons path) values and the their temporal behaviour. So, the $0.5$ ps short laser pulse created a quasi-stationary several-hundreds magnetic field structure with a characteristic decay time on $100$ ps level.

\section{Modelling of the electromagnetic structure formation}

Interaction of a laser with a curved "snail"-type target, leading to formation of a strongly magnetized plasma inside the target, needs focusing of the laser beam on the internal surface of the target entrance, see Fig.~\ref{fig_PIC_0}. The laser pulse comes at grazing incidence, which actually allows to transform the initial short pulse to a long-living electromagnetic structure, and the surface electrons are efficiently accelerated along the beam direction. At the same time, in the dense target volume,
where the laser field can not penetrate, a return electron current starts to increase to compensate a positive charge in the interaction region. Both of these electron currents, as we discuss below, create strong magnetic fields, which appear to be frozen into expanding plasma and later form a magnetized plasmoid in the target volume \cite{Korneev-pre15, ehret_kilotesla_2022}. Laser pulse follows the target geometry and this process continues all over the surface. In this interaction, a part of the most energetic electrons, which are not magnetized, escape \cite{Korneev-arxiv17} and create a positive charge of the whole target.  


\subsection{Simulations for the experimental parameters}

To model the process of laser-target interaction, we use open source Particle-In-Cell code Smilei~\cite{Derouillat2018}. For the 2D simulations, we consider a target with a realistic geometry, shown in Fig.~\ref{fig_PIC_0}. In simulations, target is made from copper with a small fraction of hydrogen, aiming to take into account a thin surface contamination layer. Simulations with the pure copper targets were also performed and showed very similar results. Laser parameters correspond to the PHELIX laser at GSI, with the beam divided into two equal parts, so that the laser energy in each is 50 J.
In these simulations, initial electron density in the target is $n_0=10^{22}$ cm$^{-3}$, copper and hydrogen ions constitute 95\% and 5\%, respectively. Spatial and temporal resolutions are $80$ nm and $0.17$ fs, respectively. Each cell has 100 electrons and 20 ions of each kind, randomly distributed in the target with initial zero temperature. Boundary conditions are open for both fields and particles. Laser beam with a 20 $\mu$m waist was focused almost with a grazing incidence on the target surface at the entrance as shown in Fig.~\ref{fig_PIC_0} at $\tau=3.49$ ps. Most of the simulations were performed till the time moment of about 20 ps, when the magnetic structure is already formed. One simulation for the S9 target was longer, up to 50 ps to demonstrate the slow hydrodynamic nature of the formed structure.


\subsection{Formation of the electromagnetic structure in a cavity}

\begin{figure}
    \centering
    \includegraphics[width=0.85\linewidth]{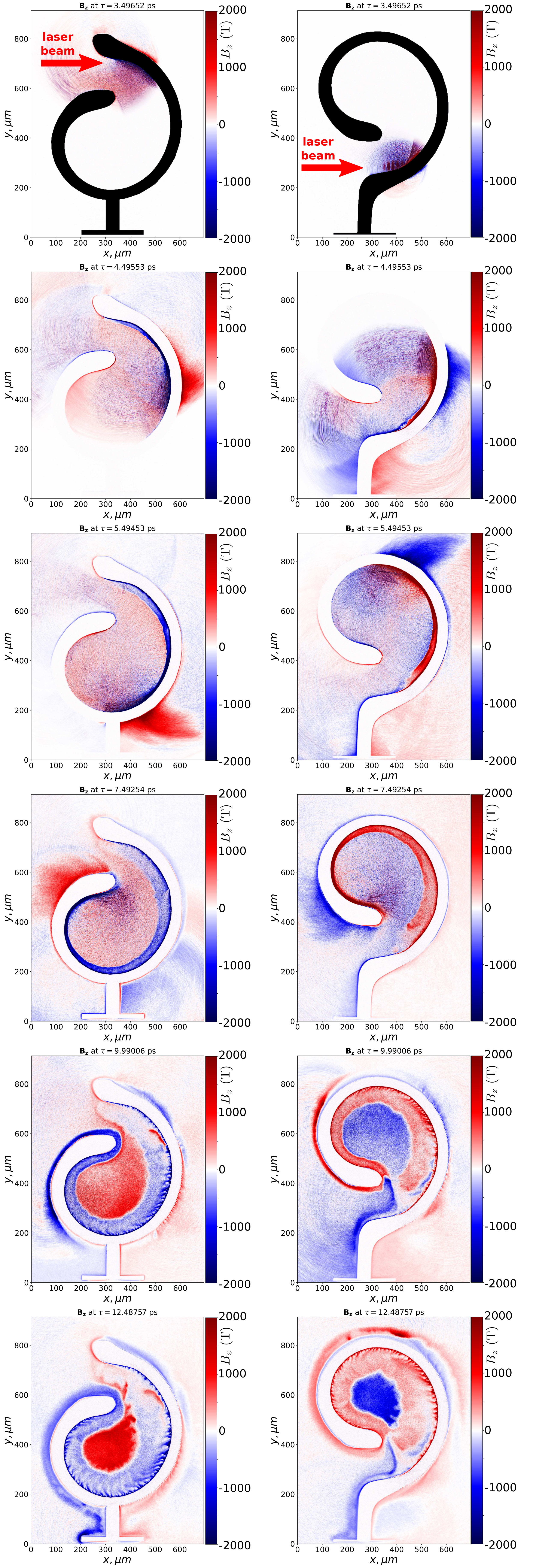}
    \caption{Initial irradiation geometry for S6 (left column) and S9 (right column) targets, and magnetic field for a sequence of time moments.}
    \label{fig_PIC_0}
\end{figure}

The interaction process looks very similar for both S6 and S9 targets. When the laser pulse comes into the target, it reflects continuously along the target surface, generating accelerated electrons predominantly in the direction of its propagation. We observe, that the propagating laser pulse is tightly bound to the surface and propagates with the discharge pulse, see Fig.~\ref{fig_PIC_0}. The target which is considered in this work is larger and more flattish than that considered in the original work \cite{Korneev-pre15} and in the experimental study performed within the same experimental campaign, but with a different target geometry \cite{ehret_kilotesla_2022}. This geometry allows to form a more symmetric structure and to confine the propagating laser pulse near the surface. 
Due to a grazing propagation all over the propagation distance, the laser pulse efficiently accelerates electrons, and the current of these electrons appears to be very distinct. As a result, outside the dense plasma region, at a distance from the surface, a positive magnetic field region is formed for the S6 target (and negative for the S9 target), while in the more dense plasma close to the surface, the direction of the field is opposite and corresponds to the closed circuit loop formed by the accelerated electrons and the return current, similarly to the previous studies. 

The temporal evolution of the magnetized plasma structure formation is presented in Fig.~\ref{fig_PIC_0}, for the S6 target in the left column and for the S9 target in the right one. We show the full magnetic field in the simulation box, the laser one and the field in the plasma. The first shown time moment $\tau\approx3.5$ ps corresponds to the start of the interaction, when the laser beam just touched the target. In this panel, the target shape and the laser beam direction is explicitly shown over the distribution of the magnetic field.

Later, at times $\tau\approx4.5, 5.5, 7.5$ ps, the laser pulse propagates with the discharge pulse along the internal surface of the target. The discharge pulse also partially covers the outer surface. The directions of the laser and the discharge pulses are opposite for the S6 and S9 targets, consequently, the formed magnetic fields have opposite directions, although qualitatively the distributions of the fields are very similar.  

\begin{figure}
    \centering
    \includegraphics[width=0.85\linewidth]{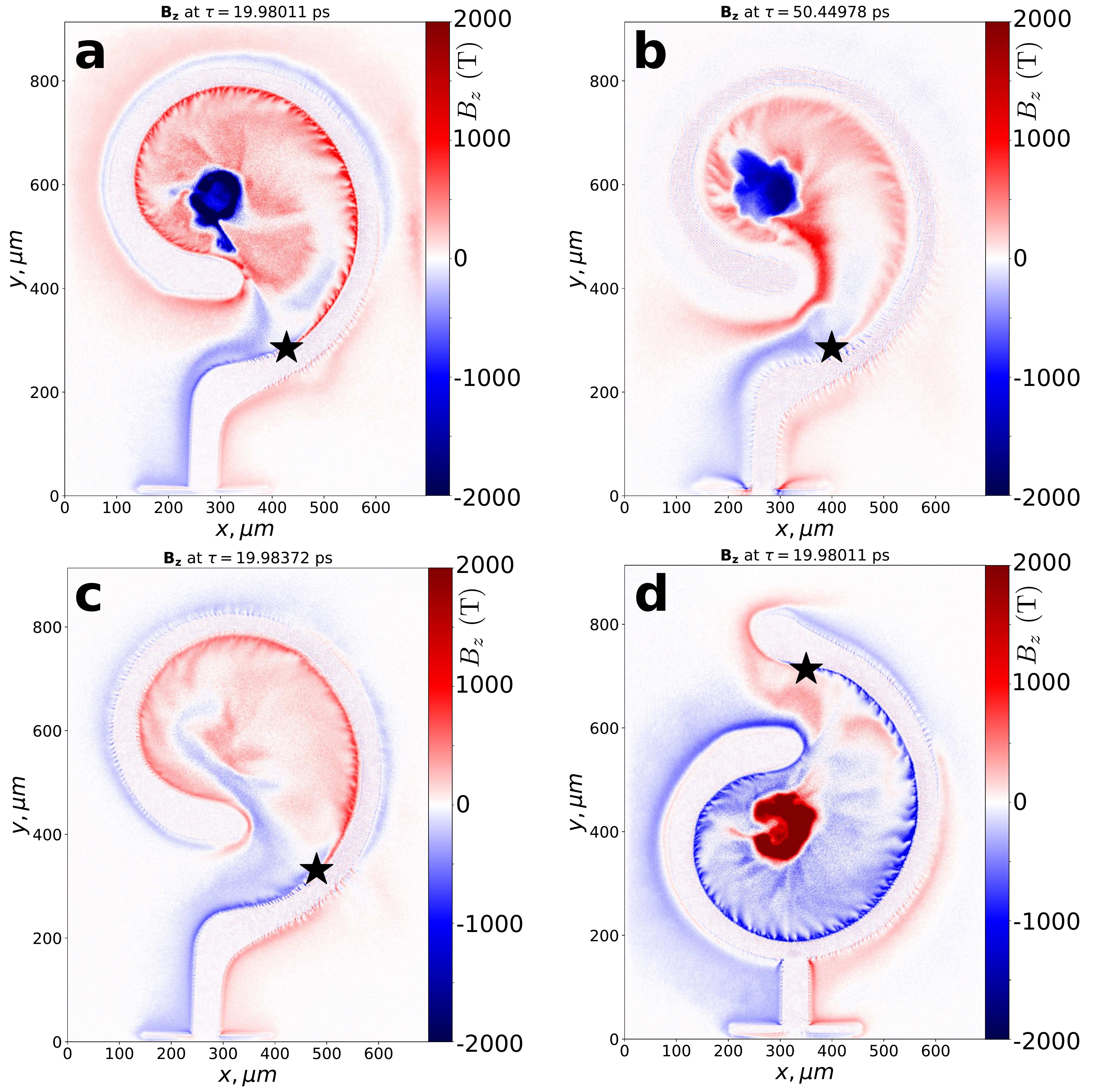}
    \caption{Magnetic structure at late times for \textbf{a}: S9 target at $\approx 20$ ps, \textbf{b}: S9 target, at $\approx 50$ ps, \textbf{c}: S9 target at $\approx 20$ ps for different focusing compared to the case shown in panel \textbf{a}, \textbf{d}: S6 target at $\approx 20$ ps. Stars on all the panels denote the focusing point.}
    \label{fig_PIC_1}
\end{figure}

At late times of about $\tau\approx10$ ps, both the discharge and the laser pulses are not evidently observed in the simulation results, see Fig.~\ref{fig_PIC_0}. Instead, a coaxial magnetized structure in the hot plasma, expanding from the internal surface, is formed. The central part of this structure has a much less density than the ring-shaped plasma closer to the surface, which contains both electrons and ions. Thus the central region is easily compressed, as seen in Fig.~\ref{fig_PIC_0} at $\tau\approx12.5$ ps. 
The later evolution proceeds on the hydrodynamical time scale to a quasi-equilibrium distribution of the magnetized plasma with a highly magnetized spot near the center, as shown in Fig.~\ref{fig_PIC_1}, see panel \textbf{a} for $\tau\approx20$ ps and panel \textbf{b} for $\tau\approx50.5$ ps.

\begin{figure}
    \centering
    \includegraphics[width=0.95\linewidth]{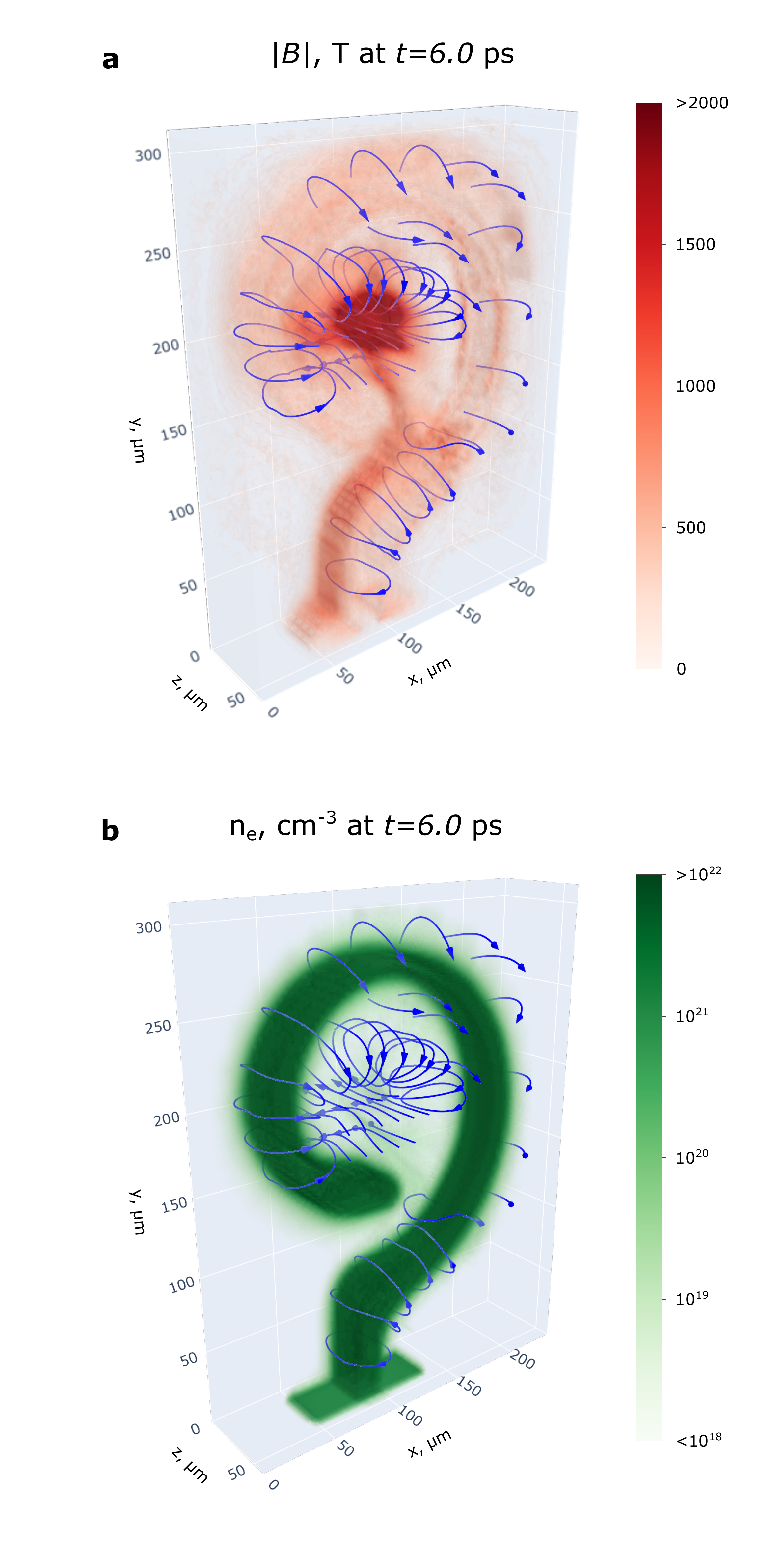}
    \caption{\textbf{a}: The magnetic field strength (color) and directions (lines with the arrows), \textbf{b}: the electron density and the magnetic field directions (lines with the arrows), obtained in the 3D PIC simulation for a reduced target.}
    \label{fig:3D_B_and_ne_with_field_lines}
\end{figure}

As mentioned above, the formation of the quasi-coaxial magnetic field distribution in the plasma of the target is related to the fine grazing conditions. In Fig.~\ref{fig_PIC_1} symbol $\bigstar$ shows the position of the laser beam focusing. Panels \textbf{a} and \textbf{b} show the magnetic field distribution formed with the optimal focusing position for a grazing incidence for the S9 target, panel \textbf{d} shows the same distribution for the S6 target. In these situations, a highly magnetized central region is surrounded by a lower and oppositely magnetized ring-shape plasma. Fig.~\ref{fig_PIC_1}~\textbf{c} represents the magnetic field structure at late time $\tau\approx20$ ps, formed by a laser beam, focused slightly upper, than it was in panels \textbf{a} and \textbf{b}, for the same S9 target. In this case, the reflection of the laser beam is not perfectly grazing, which results in qualitative changes in the magnetic field distribution. It is now more homogeneous, although with a small part of an oppositely magnetized plasma. 

To support the reasonableness of the 2D simulations, we performed an exemplary 3D PIC simulation with a reduced target size. In this simulation, the target was smaller $\sim 3$ times, the material and the number of particles were the same as in the 2D simulations, the spatial and the temporal resolutions were $10$ nm and $0.21$ fs respectively. The result for the magnetic field distribution and the electron density at $6.0$ ps are shown in Fig.~\ref{fig:3D_B_and_ne_with_field_lines}, panel \textbf{a} and \textbf{b}, respectively. It is seen, that the magnetic field distribution has a co-axial structure, with a maximum of negative field in the central region, reproducing the results of the 2D simulations. The field lines turn over the target, and then in the center change their direction, the plasma appears strongly confined near the target for the considered times. This supports the use of numerical 2D model for the general analysis presented above.  

\begin{figure}
    \centering
    \includegraphics[width=0.95\linewidth]{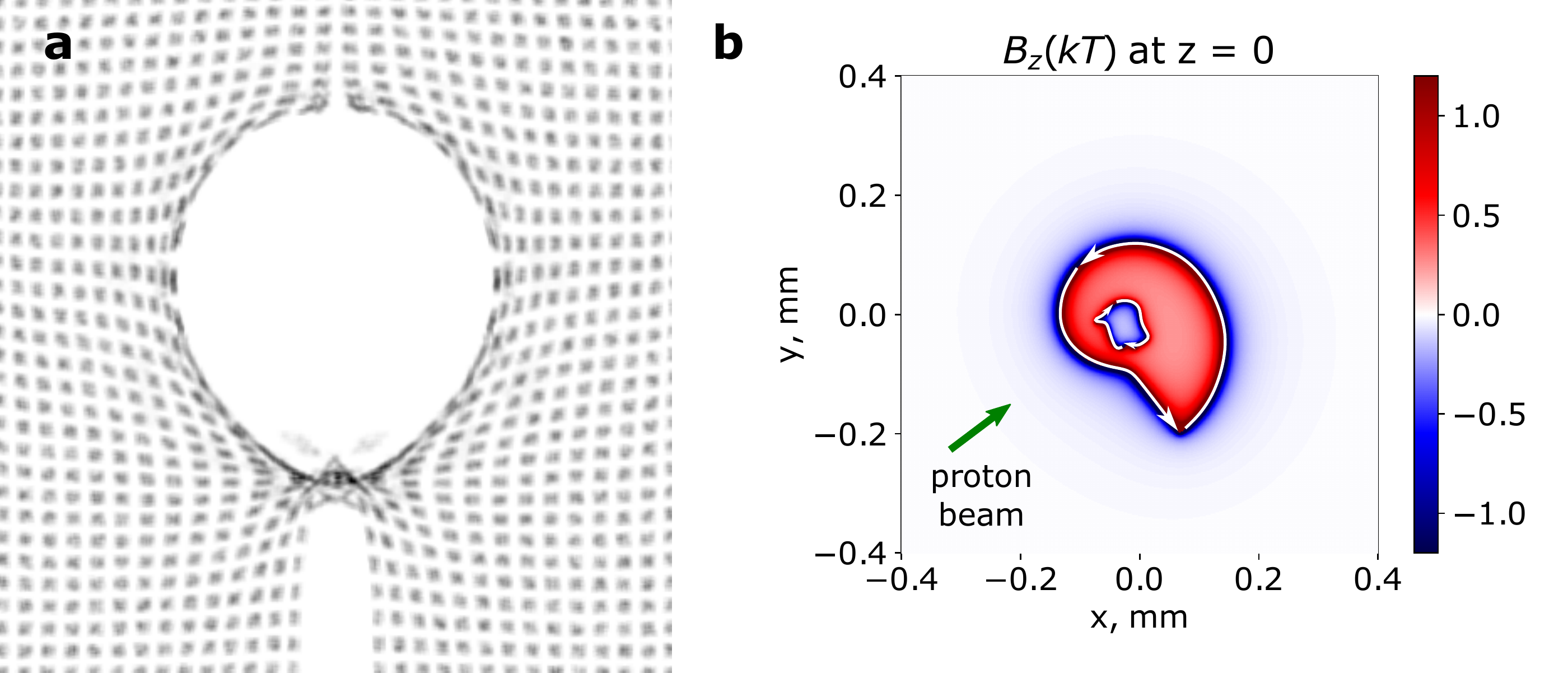}
    \caption{\textbf{a}: Synthetic radiograph, obtained for a more complex magnetic field profile, similar to that observed in PIC simulations. \textbf{b}:~The field profile, produced by two loops with different electric currents, which is parametrically adjusted to provide the highest correlation of the resultant synthetic image with the experimental one.}
    \label{fig:bipolar_B_field_prof_and_radiograph}
\end{figure}

According to the PIC simulation results, the magnetic field structure is more complicated in the case when the grazing conditions are suitable for binding the laser beam along the surface. Without a tomography probing, it is difficult to resolve the internal field structure in the experiment, and would be rather speculative to make certain conclusions on it. It is however possible to reliably define the integral field characteristics, like the average magnetic field value. To demonstrate this, a set of synthetic radiograph images for diagnostic protons propagating in a more complex magnetic field structure was built for different values of the target potential and the magnetic field values, assuming the field structure in the target volume is similar to the calculated distribution shown in Fig.~\ref{fig_PIC_1}. For that, two current loops with the shapes taken from the theoretical results were introduced. The currents in each loop, as well as the electric potential of the target, were adjusted to provide the highest correlation of the corresponding synthetic image with the experimental one. Ballistic simulation results, performed with such field structure, with a good degree of similarity reproduce both the experimental images and those, obtained in the frame of the "single loop current" model, see Fig.~\ref{fig:bipolar_B_field_prof_and_radiograph} \textbf{a}. Although without proton tomography, which involves probing the target from multiple angles, it is difficult to reliably determine the real field structure inside the cavity, for some applications, e.g. particle guiding considered in this work, both a simple uniform and a more complex co-axial magnetic field structures may appear suitable.

\section{Possible application to collimation of energetic particle flows}
As it is shown above, the curved S9 and S6 targets driven by a picosecond petawatt-class laser driver, effectively generate the electro-magnetic field structure in the plasma, which fills the target interior after heating by the laser pulse. We observe that in simulations, the field structure is sensitive to the interaction conditions. With a fine grazing reflection along the internal surface, an island of a higher magnetic field is formed near the target center, while for larger incident angles at the first interaction, the structure of the magnetic field is almost homogeneous. In experiment, it is difficult to resolve the internal field structure, but with a reasonable accuracy, using a cross-correlation analysis, an average magnetic field value in the target interior can be determined. 

\begin{figure}
    \centering
    \includegraphics[width = 0.95\linewidth]{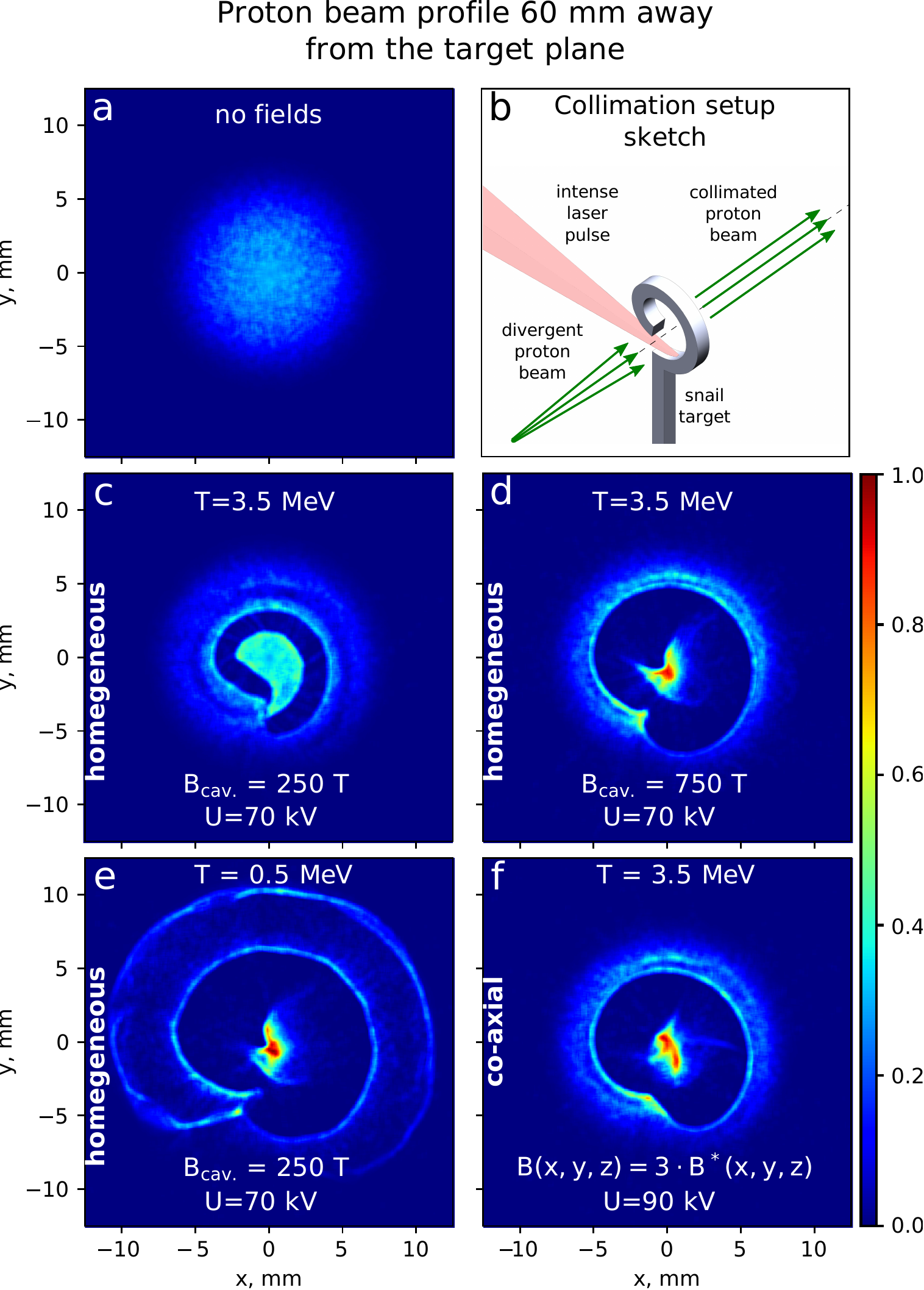}
    \caption{\textbf{a}:~Initial profile of a proton beam with a divergence half angle of $5^{\circ}$. \textbf{b}:~A sketch of a possible setup for collimation of proton beams with electromagnetic fields, created by the S9 target. \textbf{c}:~Beam profile for $3.5$~MeV protons passing through the S9 target, charged to $70$~kV, with the simple homogeneous magnetic field structure, shown in Fig.~\ref{fig:radiographs_and_MF_structure}~\textbf{b} and the average B-field value of $250$~T. \textbf{d}:~Beam profile for $3.5$~MeV protons passing through the S9 target, charged to $70$~kV, with the simple homogeneous magnetic field structure, shown in Fig.~\ref{fig:radiographs_and_MF_structure}~\textbf{b}, scaled $3$ times, so that the average B-field value is 750~T. \textbf{e}:~Beam profile for $0.5$~MeV protons passing through the S9 target, charged to $70$~kV, with the simple homogeneous magnetic field structure and the average B-field value of $250$~T. \textbf{f}:~Beam profile for $3.5$~MeV protons passing through the S9 target, charged to $90$~kV, with the complex coaxial magnetic field structure, shown in Fig.~\ref{fig:bipolar_B_field_prof_and_radiograph}~\textbf{b}, additionally scaled in magnitude by a factor of~3. Protons come from a point source $\approx2.9$~mm away from the target centre. The 'detector' plane is $60$~mm away from the target plane for all images. The images are normalized so that 1.0 corresponds to the maximum number of protons per unit area for the most highly-collimated beam.}
    \label{fig:collimation}
\end{figure}

Now, we discuss the possibility of using those optically generated electromagnetic field structures for guiding of high energy charged particle beams, considering as two examples here MeV protons and multi-MeV electrons. A sketch of the proposed setup for particle bunch collimation with one of the studied targets is illustrated in Fig.~\ref{fig:collimation} \textbf{b}. In this example, the target must be oriented so that its axis coincides with the direction of a divergent proton beam. Trajectories of protons with non-zero transverse velocity components in such a geometry will take the form of helical spirals in the region of high magnetic field - the particles will revolve around the magnetic field axis, and the particle beam may be effectively focused, forming a collimated flow. This requires certain conditions, as shown below, both for electrostatic and magnetic components of the electromagnetic field structure around the magnetized target. Surprisingly, the electrostatic field additionally enhances the collimation effect in some situations, when the positively charged protons slightly repel from positively charged quasi-ring target inwards.

In order to study proton collimation in the electromagnetic field structures, generated in snail-type targets, we employ the same test particle method that we used to analyse the experimental data. In the simulations, protons come from a point source, located $\approx2.9$~mm away from the target centre. The beam has a supergaussian angular profile with FWHM of $10^{\circ}$. The detector plane is in $60$~mm distance from the target plane. Initial beam profile for the case, when no external fields are applied, is shown in Fig.~\ref{fig:collimation}~\textbf{a}. First, we consider the simple homogeneous field geometry, see Fig.~\ref{fig:radiographs_and_MF_structure}~\textbf{c} and $3.5$~MeV proton beam. In this case, if the values of magnetic and electric fields are close to those obtained from the experimental data analysis, the fields considerably change the proton concentration per unit area for particles that pass the target area, but the collimation is not perfect, see Fig.~\ref{fig:collimation}~\textbf{c}. With these fields, a better collimation may be obtained for lower proton energies, see Fig.~\ref{fig:collimation}~\textbf{e} for protons with the energy of $0.5$~MeV. For the considered $3.5$~MeV protons a good collimation can be achieved, if the values of magnetic field are scaled by a factor of~$3$, so that the average magnetic field inside the cavity reaches $750$~T. Such an increase of the fields values is possible either with the increase of the invested laser energy, or, for the same laser energy, with the decrease of the target size~\cite{ehret_kilotesla_2022} which may require a change of the setup geometry. As can be seen in Fig.~\ref{fig:collimation}~\textbf{d}, under the action of such increased fields, protons strongly redistribute and a significant fraction of them is focused in the central region, forming a bright spot with $\approx3$ times higher proton concentration at the detector plane. Creating average magnetic fields of $750$~T with the same target geometry requires $\sim9$ times higher laser energies, that is about $450$~J. Such values in picosecond regime are well within reach for modern high power laser systems, such as, for example, Z-petawatt laser at Sandia National Laboratories in the USA, PETAL laser in France, or LFEX facility in Japan.

A further criterion for the applicability of the produced electromagnetic fields is their rise and decay over time. Achromatic electromagnetic lenses are one application that can be realized if the focal length remains unchanged for all projectile energies. Such lenses are important for beam collimation over a large spectral bandwidth, e.g. for bunching with high frequency (HF) cavities~\cite{Ja2019}. For heuristic estimates we regard the focal length of a conductive ring of radius $R$ at potential $U$ which holds a coiling current inducing a magnetic field $B$, for details see appendix~\ref{sec:achromat}. Following a thin lens approximation, the focal length $f$ is constant when $\alpha U/R + \beta R B^2 = f(t-t_0)^{-2}$. The proportionality factors $\alpha \approx 0.07 q/m$ and $\beta \approx 0.29 (q/m)^2$ result from the ring geometry of the fields and the projectile charge to mass ratio. Perfect lensing requires a slow decay $B \propto 1/t$ and a faster decay $U \propto 1/t^2$ -- a trend that is qualitatively in agreement with the experimental data, see their respective evolution in {Fig.~\ref{fig:correlation_peak_and_B(t)}~\textbf{c}}. Here, the experimental values are suited to maintain a quasi-constant focal length of $2.9$~mm for protons with energies ranging from $360$~keV to $380$~keV launched $t_0 = 331$~ps before driving the fields. It is possible to shift the spectral range, either by scaling the field amplitudes or by modification of the source distance. Note, that the potential can be decreased without altering the magnetic field strength by deploying more massive targets with the same internal geometry. Consider for example parametric changes that may appear beneficial to reach applicability of the platform at the same focal length not taking into account the target potential. (I) increasing the magnetic field by $45$~\% yields acceptance of protons with energies ranging from $620$~keV to $670$~keV launched $t_0 = 246$~ps before driving the fields. An optimization of the platform towards larger spectral acceptance may become possible by tailoring of the characteristic decay times of fields, e.g. by changing the target material. With tailored magnetic field decay the temporal range of measured data allowed for reaching lower energies down to $550$~keV, the energy range interesting for proton-Boron fusion experiments. (II) scaling up the fields by a factor of $7$ allows to maintain the constant focal length for protons in a larger energy bin from $7$~MeV to $17$~MeV when launched $40$~ps ahead of the field drive, the energy range interesting for plasma accelerators aiming at medical applications.

Now, consider a more complex co-axial magnetic field structure, observed in PIC simulation, with electromagnetic field parameters deduced from the experimental data, see Fig.~\ref{fig:bipolar_B_field_prof_and_radiograph}~\textbf{b}. It appears, that this field profile may be suited for particle guiding as well as the homogeneous one. According to the results of ballistic calculations, non-scaled values of coaxial magnetic fields, deduced from the experimental radiograph are again sufficient to collimate $0.5$~MeV protons. As well, more energetic $4.5$~MeV protons can be guided if the coaxial magnetic field is increased by a factor of $\sim3$, see Fig.~\ref{fig:collimation}~\textbf{f}. Taking into account the fact that electric fields also play a significant role in particle collimation, the requirements on laser energy for collimation of $\approx 3.5$~MeV protons with both the homogeneous and the coaxial field structures may appear to be less, since the potential in this case will be higher than $70-90$~kV, deduced from the experimental data obtained for $50$~J laser pulses.

\begin{figure}
    \centering
    \includegraphics[width = 0.95 \linewidth]{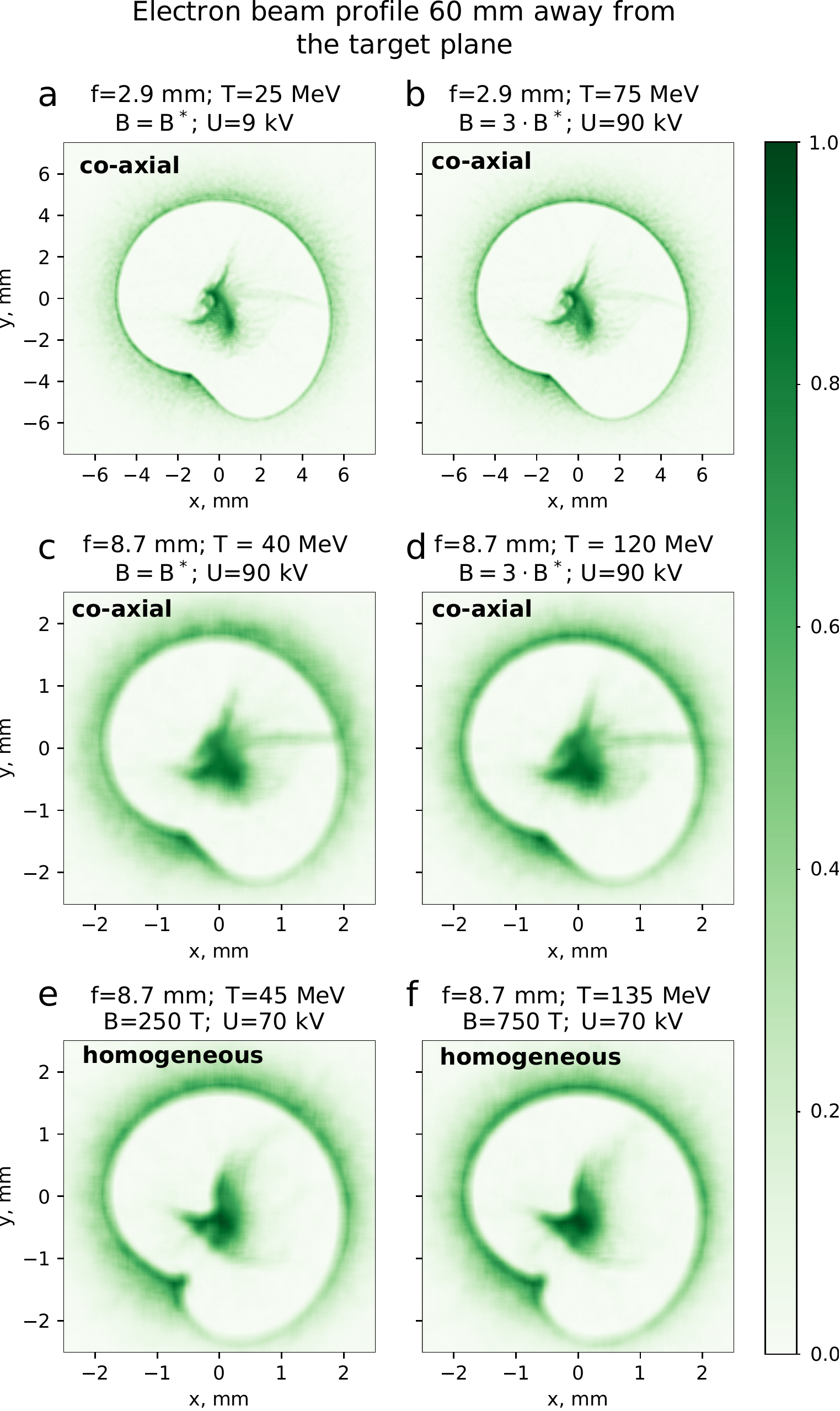}
    \caption{\textbf{a}:~Beam profile for 25~MeV electrons, coming from a point source $\approx$2.9~mm away from the target plane and passing through the S9 target, charged to 90~kV, with the complex coaxial magnetic field structure, shown in Fig.~\ref{fig:bipolar_B_field_prof_and_radiograph}, \textbf{b}. \textbf{b}:~Beam profile for 75~MeV electrons, coming from a point source $\approx 2.9$~mm away from the target plane and passing through the S9 target, charged to 90~kV, with the complex coaxial magnetic field structure, additionally scaled in magnitude by a factor of~3. \textbf{c}:~Beam profile for 40~MeV electrons, coming from a point source $\approx 8.7$~mm away from the target plane and passing through the S9 target, charged to 90~kV, with the complex coaxial magnetic field structure. \textbf{d}:~Beam profile for 120~MeV electrons, coming from a point source $\approx 8.7$~mm away from the target plane and passing through the S9 target, charged to 90~kV, with the complex coaxial magnetic field structure, additionally scaled in magnitude by a factor of~3. \textbf{e}:~Beam profile for 45~MeV electrons, coming from a point source $\approx 8.7$~mm away from the target plane and passing through the S9 target, charged to 70~kV, with the simple homogeneous field structure and the average B-field value of 250~T. \textbf{f}:~Beam profile for 135~MeV electrons, coming from a point source $\approx 8.7$~mm away from the target plane and passing through the S9 target, charged to 70~kV, with the simple homogeneous magnetic field structure and the average B-field value of 750~T. The 'detector' plane is 60~mm away from the target plane for all images. The images are normalized so that 1.0 corresponds to the maximum number of electrons per unit area for the most highly-collimated beam.}
    \label{fig:collimation_eon}
\end{figure}

Additionally, we performed ballistic simulations with relativistic electrons to estimate in what energy range they can be collimated with the proposed setup. In the case of the complex coaxial magnetic field structure with the parameters deduced from the experimental data, already 25~MeV electrons can be efficiently collimated (see Fig.~\ref{fig:collimation_eon}, \textbf{a}). Scaling this magnetic field by a factor of 3 enables the collimation of 75~MeV electrons (see Fig.~\ref{fig:collimation_eon}, \textbf{b}). 
It should be noted that for electrons the electric field, acting on the particles when they travel trough the cavity of the positively charged target, creates a force that is directed outwards rather than inwards, as it is in the case of protons, and thus, it leads to some decollimation. 

An alternative way of increasing the energy of the particles that can be guided with the proposed setup is moving the particle source away from the target plane rather than increasing the magnitude of the magnetic field. It is based on the fact that transverse velocity components for particles that come from a more distant source and pass through the target cavity in this case are lower, and thus less deflection is required to make their trajectories parallel to the target axis. As can be seen from Fig.~\ref{fig:collimation_eon}~\textbf{c}, increasing the distance from the particle source to the target by 3~times allows the collimation of 40~MeV electrons with the magnetic field values deduced from the experimental data, while the field scaled in magnitude by $3$ times enables the guiding of 120~MeV electrons, see Fig.~\ref{fig:collimation_eon}~\textbf{d}. Coming back to the homogeneous field structure, it has similar collimation characteristics for both short and long focal distances. It results in slightly higher collimated particle energies, as well as slightly higher particle concentration in the collimated spot, although the shape of the spot is more complex than with the coaxial magnetic field structure, see Fig.~\ref{fig:collimation_eon}~\textbf{e}~and~\textbf{f}. This approach of increasing the energy range of the guided projectiles may be well suited for particle acceleration schemes with initially low beam divergence, so that the source can be moved further away from the magnetic lens while still maintaining relatively high number of particles that are initially sent through the high magnetic field region. In addition, multi-stage collimation schemes may be considered if several laser beams are available, this situation would be considered elsewhere.         

\section{Conclusion}

In this work, the curved snail-type targets, which we called "S6" and "S9", depending on their helicity, were shown to be effective optical generators of strong magnetic fields with use of picosecond petawatt-class laser facilities. The generation of the high magnetic field of about several hundreds of tesla with only a $50$~J laser pulse was observed both in simulations and experiment. The considered picosecond regime of a driver laser pulse is consistent with widely studied schemes for generation of highly energetic proton and ion beams, which makes the setup more easily adjustable, e.g. more easily synchronized. The considered targets possess a relatively large "useful" surface, where the particles may be focused, of about $0.5$~mm diameter. 
It is found that the fine reflection conditions on the internal target surface result in the formation of a "coaxial" magnetized structure in the target volume, which is filled with a hot magnetized plasma and evolve after formation on a hydrodynamical time scale of about a hundred of picoseconds. The coaxial structure appears as suitable for energetic particle guiding as the homogeneous one, providing similar collimation efficiency.  
The proposed setup may appear a perspective proton and ion post-generation guiding line, providing good collimation efficiency for multi-MeV protons and multi-hundreds-MeV electrons with modern and under-construction picosecond petawatt laser facilities.

\begin{acknowledgments}
The results presented here are based on the experiment P136, which was performed at the PHELIX facility at the GSI Helmholtzzentrum f\"{u}r Schwerionenforschung, Darmstadt (Germany) in the frame of FAIR Phase-0. The authors are grateful to the PHELIX team for support in the data collection campaign that took place before Feb. 24, 2022. 
The work was partially supported by the project \# FSWU-2023-0070 (Ministry of Science and Higher Education of the Russian Federation). This work was also granted access to the HPC resources of CINES under allocations Nos. A0020510052 and A0030506129 made by GENCI (Grand Equipment National de Calcul Intensif) and access to MEPhI HPC resources.
\end{acknowledgments}

\appendix

\section{}\label{sec:achromat}

A thin magnetic lens with geometry of a perfect coil with radius $R$ has a focal length of $f_\mathrm{B} = 64mK/3\pi q^2 R B^2$ for non relativistic projectiles of kinetic energy $K$, mass $m$ and charge $q$, where $B$ denotes the field strength in the centre of the coil \cite{RE2008}. Here, particles are supposed to propagate close to the coil axis and deflection angles are presumed to be small. A quasi-neural particle beam composed of ions and slow co-propagating electrons will experience a stripping of negative charge due to the strong effect of the Lorentz force on electrons. Stripping electrons ahead of the coil may affect focusing of the magnetic lens, even though such space charge effects are relatively low~\cite{Co2004} in typical laser accelerated beams issued via TNSA. Where stripping of electrons does not degrade magnetic lensing, it enables for electric lensing in the configuration of a positively charged ring. For a beam propagating along the coil symmetry axis $z$ with charge $q(z)$ that is accelerated by the radial component of the electric field $E_r(z)$, the integral of the Lorentz force $\int_{-\infty}^{+\infty} q E_r \text{~d}z$ vanishes. Stripping results in an asymmetric $q(z)$ that allows for a net deflection towards the axis $\alpha_r \propto \delta r$ with $\delta r / R \ll 1$. The later proportionality is a necessary condition for the formation of a focal spot. For further simplification and estimates, one assumes now that the stripping occurs instantaneously at the position prior to the coil plane where the radial electric field swings from positive to negative values. Then the focal length of the electric lens is $f_\mathrm{U} = 200KR/7qU$, where $U$ denotes the uniform potential of the coil. In the limit of thin lenses, the focal length of a combined electric and magnetic lens calculates to $f = (1/f_\mathrm{B} + 1/f_\mathrm{U})^{-1}$.

\bibliography{local}

\end{document}